\documentclass[11pt]{article}

\usepackage{graphicx}
\usepackage[reqno,tbtags]{amsmath}
\usepackage{multirow}
\usepackage{cite}

\allowdisplaybreaks

\newcommand\pubnumber{ PITHA 08/20  \\
                       SFB/CPP-08-62  \\
                       TTP08-38 }

\newcommand\pubdate{\today}

\def\csuma{Institut f\"ur Theoretische Physik E, RWTH Aachen University,\\
           52056 Aachen, Germany}
\def\csumb{Dipartimento di Fisica Teorica, Universit\`a di Torino, Italy\\
           INFN, Sezione di Torino, Italy}
\def\csumc{Physics Department, Brookhaven National Laboratory,\\
           Upton, NY 11973, USA}
\def\csumd{Institut f\"ur Theoretische Teilchenphysik, Universit\"at Karlsruhe,\\
           76128 Karlsruhe, Germany}

\def\Title#1{\begin{center}{\Large \bf #1}\end{center}}

\def\Author#1{\begin{center}{\sc #1}\end{center}}

\def\Address#1{\begin{center}{\it #1}\end{center}}

\newcommand\pubblock{\rightline{\begin{tabular}{l}\pubnumber\\
                     \pubdate\\\end{tabular}}}

\newenvironment{Abstract}{\begin{quotation}}{\end{quotation}}

\def\Acknowledgments{\bigskip\bigskip\begin{center}
                     \large\bf Acknowledgments\end{center}}

\def\email#1{\footnote{#1}}

\makeatletter

\def\section{\@startsection{section}{0}{\z@}{5.5ex plus .5ex minus
                     1.5ex}{2.3ex plus .2ex}{\large\bf}}

\def\subsection{\@startsection{subsection}{1}{\z@}{3.5ex plus .5ex minus
                          1.5ex}{1.3ex plus .2ex}{\normalsize\bf}}

\def\subsubsection{\@startsection{subsubsection}{2}{\z@}{-3.5ex plus
                                      -1ex minus  -.2ex}{2.3ex plus .2ex}{\normalsize\sl}}

\input Input_rosetta.sty

\newcommand{\Fsc}{\mu^2_{\ssF}}
\newcommand{\Rsc}{\mu^2_{\ssR}}

\begin{document}

\begin{titlepage}
  \pubblock
  \vfill
  \def\thefootnote{\fnsymbol{footnote}}
  \Title{NLO Electroweak Corrections to Higgs Boson\\[3mm]
         Production at Hadron Colliders 
  \footnote[9]{Work supported by MIUR under contract 2001023713$\_$006, 
               by the European Community's Marie Curie Research 
               Training Network {\it Tools and Precision Calculations for Physics Discoveries 
               at Colliders} under contract MRTN-CT-2006-035505, by the U.S. 
               Department of Energy under contract No. DE-AC02-98CH10886 and
               by the Deutsche Forschungsgemeinschaft through Sonderforschungsbereich/Transregio 9 
               {\it Computergest\"utzte Theoretische Teilchenphysik}.
               We thank the Galileo Galilei institute for Theoretical Physics
               for hospitality and the INFN for partial support during the completion of
               this work.}}
\vfill
\Author{Stefano Actis       \email{actis@physik.rwth-aachen.de}}        \Address{\csuma}
\Author{Giampiero Passarino \email{giampiero@to.infn.it}}               \Address{\csumb}
\Author{Christian Sturm     \email{sturm@bnl.gov}}            \Address{\csumc}
\Author{Sandro Uccirati     \email{uccirati@particle.uni-karlsruhe.de}} \Address{\csumd}
\vfill
\begin{Abstract}
\noindent
  Results for the complete NLO electroweak corrections to Standard 
  Model Higgs production via gluon fusion are included in the total cross section  
  for hadronic collisions. Artificially large threshold effects are avoided 
  working in the complex-mass scheme. The numerical impact at LHC (Tevatron) 
  energies is explored for Higgs mass values up to $500\,$ GeV ($200\,$ GeV). 
  Assuming a complete factorization of the electroweak corrections, one finds 
  a $+\,5\, \%$ shift with respect to the NNLO QCD cross section for a Higgs mass 
  of $120\,$ GeV both at the LHC and the Tevatron. Adopting two different 
  factorization schemes for the electroweak effects, an estimate of the 
  corresponding total theoretical uncertainty is computed. 
\end{Abstract}
\vfill
\begin{center}
\end{center}
\end{titlepage}
\def\thefootnote{\arabic{footnote}}
\setcounter{footnote}{0}
\clearpage
\setcounter{page}{1}
\section{Introduction}
Gluon fusion is the main production channel for the Standard Model 
Higgs boson at hadron colliders. Unsurprisingly, radiative corrections 
have been thoroughly investigated in the past years; in particular, 
since next-to-leading order (NLO) QCD corrections increase the inclusive 
cross section for Higgs production at the LHC by a factor of about $1.5$ to 
$1.7$ with respect to the leading order (LO) term~\cite{Spira:1995rr}, 
there was a flurry of activity on higher order QCD effects. Recent 
reviews on the subject can be found in Ref.~\cite{Catani:2006kr}.

Electroweak effects are less understood than QCD ones. An early 
study~\cite{Djouadi:1994ge} concluded that the so-called leading NLO 
corrections to the partonic gluon-fusion cross section $\sigma(gg \to H)$, 
enhanced by the squared mass of the top quark, amount to $0.4\, \%$. The 
size of this result is due to the delicate cancellation mechanism
among various contributions described in Ref.~\cite{Djouadi:1994ge}, 
and more recent computations have shown that these corrections are not 
the dominant ones in the entire Higgs mass range.

Contributions induced by light quarks have been calculated
in Ref.~\cite{Aglietti:2004nj} and found to be extremely larger, reaching 
a maximum of about $9\, \%$ for a Higgs mass below $160\,$ GeV. The terms
of the amplitude involving the top quark have been evaluated in
Ref.~\cite{Degrassi:2004mx} by means of a Taylor expansion in the kinematic 
region below the $WW$ threshold, where they partially screen the dominant 
effect of the light quarks. The impact of the partonic results of 
Refs.~\cite{Aglietti:2004nj,Degrassi:2004mx} on the total cross section 
for Higgs production in proton-proton collisions, $\sigma(pp \to H + X)$, 
has been estimated for LHC energies in Ref.~\cite{Aglietti:2006yd}. Assuming 
a complete factorization of the electroweak effects with respect to the dominant 
soft and collinear QCD radiation, the analysis shows that the cross section 
increases in a range from $4$ to $8\, \%$  for a Higgs mass $\mh \le 160\,$ GeV.

A satisfactory understanding of NLO electroweak effects on Higgs production
at hadron colliders requires at least two additional steps: first, a detailed 
analysis of the $WW$, $ZZ$ and $t {\bar t}$ threshold regions, looking ahead to 
the possible occurrence of artificially large effects associated with the opening 
of two-particle thresholds; second, an extension of the corrections due to the 
top quark to the entire Higgs mass range. Furthermore, because of the advent of 
the LHC, an independent derivation of the results of 
Refs.~\cite{Aglietti:2004nj,Degrassi:2004mx,Aglietti:2006yd} for a light Higgs 
is certainly justified.

Improving the methods employed in Ref.~\cite{Passarino:2007fp} in the context
of the Standard Model Higgs decay to two photons, we have recently completed 
the evaluation of all NLO electroweak corrections to the gluon-fusion Higgs 
production cross section at the partonic level~\cite{newLong}, deconvoluted of 
the well-known QCD effects~\cite{Spira:1995rr}. 

We have found that the corrections enhance the production mechanism
throughout a Higgs mass range spanning from $100\,$ GeV to about $180\,$ GeV,
where light quarks dominate and the full NLO contributions to the partonic 
cross section $\sigma(gg \to H)$ reach up to $6\, \%$. For higher 
values of $\mh$, instead, the corrections become negative and light 
quarks are not dominating; a minimum of $-4\, \%$ is reached 
around the $t {\bar t}$ threshold.

In addition, we have performed a dedicated study of the behavior around 
the $WW$, $ZZ$ and $t {\bar t}$ thresholds~\cite{newShort}, showing that
unphysical singularities and large threshold effects disappear once
the complex-mass scheme of Ref.~\cite{Denner:2005fg} is applied
in a two-loop context following the strategy described in 
Ref.~\cite{Actis:2006rc}.

In this paper we present our numerical results for the inclusive Higgs production 
cross section in hadronic collisions, including our own evaluation of the NLO 
electroweak corrections in addition to the next-to-next-to-leading order 
(NNLO) QCD effects~\cite{Harlander:2000mg,Anastasiou:2002yz,Kilgore:2002sk}.
Moreover, we provide an estimate of the residual theoretical uncertainty and 
perform a comparison with the impact of soft-gluon resummation
at next-to-next-to-leading logarithmic (NNLL) level~\cite{Catani:2003zt}.

Our detailed numerical study is motivated by the observation that a
typical size for the NLO electroweak corrections at the partonic level 
is $5\, \%$ for $\mh=120\,$ GeV; this value matches the theoretical 
uncertainty associated with uncalculated higher order QCD corrections, 
estimated to be $5\, \%$ at the LHC and $7\, \%$ at the 
Tevatron~\cite{Moch:2005ky}.
\section{Inclusion of the NLO electroweak corrections}
\label{sec:inclu}
The inclusive cross section for the production of the Standard Model
Higgs boson in hadronic collisions can be written as
  \bqa 
  \label{eq:CShad}
    \sigma \lpar s,\mhs \rpar &=&
    \sum_{i,j} \, \int_0^1 \! dx_1  \int_0^1 \! dx_2 \,\,
    f_{i / h_1}\lpar x_1,\Fsc \rpar \, 
    f_{j / h_2}\lpar x_2,\Fsc \rpar \,
  \times \nl {}&\times&
    \int_0^1 \! dz \, \delta \lpar z -\frac{\mhs}{s\, x_1 x_2} \rpar 
    \,z\, \sigma^{(0)}\, 
    G_{ij}\lpar z;\alpha_{\ssS}(\Rsc),\mhs/\Rsc; \mhs/ \Fsc \rpar,
  \eqa
where $\sqrt{s}$ is the center-of-mass energy and $\mu_{\ssF}$ and 
$\mu_{\ssR}$ stand for factorization and renormalization scales.

In \eqn{eq:CShad} the partonic cross section for the sub-process $ij\to H+X$, 
with $i(j) = g, q_f, {\bar q}_f$, has been convoluted with the parton 
densities $f_{a/h_b}$ for the colliding hadrons $h_1$ and $h_2$. The Born 
factor $\sigma^{(0)}$ reads
  \bq
  \label{eq:oth}
    \sigma^{(0)} =  \frac{\gf}{288\sqrt{2}\pi} 
                 \left| \frac{3}{2} \sum_q  \frac{1}{\tau_q}  
                   \left[ 1+\left(1-\frac{1}{\tau_q}\right) 
                     f(\tau_q)  \right]  \right|^2,
  \eq
where $\gf$ is the Fermi-coupling constant, $\tau_q= \mhs / (4 M_q^2)$  and $M_q$ is
the conventional on-shell mass of the top or bottom quark; the function $f$ is
  \bq
    f(\tau_q)  = \left\{ \begin{array}{ll}
    \displaystyle \arcsin^2 \sqrt{\tau_q},  & \qquad \tau_q \le 1, \\
    \displaystyle - \frac{1}{4} 
    \left[ \ln \frac{1+\sqrt{1-\tau_q^{-1}}}{1-\sqrt{1-\tau_q^{-1}}} 
      - i\pi \right]^2,  & \qquad \tau_q > 1
    \end{array} \right. .
  \eq

The coefficient functions $G_{ij}$ can be computed in QCD through a 
perturbative expansion in the strong-coupling constant $\alpha_\ssS$,
  \bq
    G_{ij} \lpar z ; \alpha_{\ssS}(\Rsc) , \mhs/\Rsc ; \mhs/\Fsc \rpar =
    \alpha_{\ssS}^2(\Rsc) 
    \sum_{n=0}^{\infty} \lpar \frac{\alpha_{\ssS}(\Rsc)}{\pi}\rpar^n
    G_{ij}^{(n)}\lpar z;\mhs/\Rsc;\mhs/\Fsc \rpar,
  \eq
with a scale-independent LO contribution given by
  $
    G^{(0)}_{ij}(z) = \delta_{ig}\,\delta_{jg}\,\delta\lpar 1 - z\rpar.
  $
The NLO QCD coefficients have been computed in Ref.~\cite{Spira:1995rr}, 
keeping the exact $M_t$ and $M_b$ dependence. NNLO results have been
derived in Ref.~\cite{Harlander:2000mg} in the large $M_t$ limit (see
Ref.~\cite{Dawson:1990zj} for the NLO case); analytical 
expressions can be found in Ref.~\cite{Anastasiou:2002yz}  (an independent 
cross-check has been reported in Ref.~\cite{Kilgore:2002sk}). 
The accuracy of these fixed-order computations has been improved with soft-gluon
resummed calculations~\cite{Catani:2003zt,Moch:2005ky,Laenen:2005uz}.

The inclusion of higher order electroweak corrections in \eqn{eq:CShad} 
requires to define a factorization scheme (relevant examples on non-factorizable
effects concerning $Z$ boson decay can be found in Refs.~\cite{Fleischer:1992fq,
Czarnecki:1996ei}). 
The authors of Ref.~\cite{Aglietti:2006yd} assume that the modifications 
induced by sub-leading higher order terms starting at three loops 
are small; consistently with this assumption, they completely factorize 
QCD and electroweak corrections at the partonic level. This ansatz is certainly 
well justified for $\mh \ll \mw$, where the electroweak interaction is effectively 
point-like. 

Since our analysis spans the entire Higgs mass range, and we do not
foresee the advent of a three-loop calculation involving mixed QCD 
and electroweak effects, we adopt a more conservative approach,
and resort to the well-established LEP practice of attributing a theoretical 
error to the inclusion of NLO electroweak corrections~\cite{Bardin:1997xq}. 

We introduce two options for replacing the purely QCD-corrected partonic cross 
section in \eqn{eq:CShad} with the expression including NLO electroweak corrections:
\bei
  \item[--] CF (Complete Factorization):
    \bq \label{eq:CF}
    \sigma^{(0)}\, G_{ij} \to \sigma^{(0)}\,\lpar 1 + \delta_{\EW}\rpar\,G_{ij};
    \eq
  \item[--] PF (Partial Factorization):
    \bq \label{eq:PF}
    \sigma^{(0)}\,G_{ij} \to \sigma^{(0)}\,\Bigl[ G_{ij} + \alpha_\ssS^2(\Rsc) 
    \delta_{\EW}\,G^{(0)}_{ij}\Bigr],
    \eq
\eei
where $\delta_{\EW}$ embeds all NLO electroweak corrections to the partonic
cross section $\sigma(gg\to H)$, 
\bq
\label{eq:deltaPART}
\sigma_{\EW}=\alpha_{\ssS}^2(\Rsc)\sigma^{(0)}(1+\delta_{\EW}),
\eq
with $\sigma^{(0)}$ defined in \eqn{eq:oth}. The CF option of \eqn{eq:CF} amounts 
to an overall re-scaling of the QCD result, dressed at all orders with the NLO
electroweak correction factor $\delta_{\EW}$; the PF option of \eqn{eq:PF} is 
equivalent to add electroweak corrections to QCD ones. 

An intermediate option 
would be to fold the NLO electroweak corrections with the pure gluon-gluon NLO 
and NNLO QCD components;  in this case $\delta_{\EW}$ would be convoluted with 
$G^{(1)}_{gg}$, but not with $G^{(1)}_{gq}$ and $G^{(1)}_{q {\bar q}}$.

Next, we define the uncertainty due to uncalculated higher order QCD 
corrections according to the standard method used in Ref.~\cite{Catani:2003zt}:
we vary the renormalization and factorization scales $\mu_{\ssR}$ and 
$\mu_{\ssF}$ around the natural scale of the process $\mh$, changing their values
first simultaneously, with the bound  $\mh/2 \leq \mu_{\ssR,\ssF} \leq 2\,\mh$,
and then independently, with the additional constraint 
$\mu_{\ssR}/2 \leq \mu_{\ssF} 
\leq 2\,\mu_{\ssR}$ at fixed $\mu_\ssR$. For each value of $\mh$, the minimal (maximal) 
values of the inclusive cross section of \eqn{eq:CShad},
associated with $\mu_{\ssR, \rm min}$ and $\mu_{\ssF, \rm min}$ 
($\mu_{\ssR, \rm max}$ and $\mu_{\ssF, \rm max}$) are denoted by 
$\sigma^{\rm QCD}_{\rm min}$ ($\sigma^{\rm QCD}_{\rm max}$); their difference 
defines the QCD uncertainty band around the reference value 
$\sigma^{\rm QCD}_{\rm ref}$, obtained setting $\mu_{\ssR}=\mu_{\ssF}=\mh$.

Finally, for each set $\{\mh; \mu_{\ssR,\rm min}; \mu_{\ssF,\rm min}\}$,
$\{\mh; \mu_{\ssR}=\mh; \mu_{\ssF}=\mh\}$ and 
$\{\mh; \mu_{\ssR,\rm max}; \mu_{\ssF,\rm max}\}$ we repeat the computation of 
\eqn{eq:CShad} performing the replacements of \eqn{eq:CF} and \eqn{eq:PF}.
As a result, two new reference values are obtained for 
both CF and PF options, giving the impact of our evaluation of the NLO electroweak 
corrections.
In addition, we obtain new minimal and maximal values for the cross section,
dressed with electroweak effects, and define a new uncertainty band.

Needless to say, this procedure will only give an approximate bound on the
total uncertainty; one should indeed observe that for the considered range the LO 
and NLO QCD bands of Ref.~\cite{Spira:1995rr} do not overlap, with an NNLO band 
which is only partly contained in the NLO one.
\section{Numerical results}
For the NLO electroweak corrections we use our recent result~\cite{newLong} 
and consider a Higgs mass range spanning from $100\,$GeV to $500\,$GeV.
In this region we cross the $WW$, $ZZ$ and $t\bar{t}$ thresholds. A naive 
computation of the amplitude with conventional on-shell masses as input data 
reveals the presence of singularities at the $WW$ and $ZZ$ thresholds; 
in order to cure them, we have introduced in our computation 
complex masses~\cite{Actis:2006rc}, following the suggestion of 
Ref.~\cite{Denner:2005fg}. The behavior at the $t\bar{t}$ thresholds, 
instead, is smooth, and the on-shell mass of the top quark can be 
safely used.

In the calculation all light-fermion masses have been set to zero and 
we have defined the $W$ and $Z$ boson complex poles by
\bq
s_{j} = \mu_{j}\,\lpar \mu_{j} - i\,\gamma_{j}\rpar,
\quad 
\mu^2_{j} = M^2_j - \Gamma^2_{j},
\quad
\gamma_{j} = \Gamma_{j}\,\left( 1 - \frac{\Gamma^2_{j}}{2\,M^2_j}\right),
\label{replacement}
\eq
with $j=W,Z$. As input parameters for the numerical evaluation we have used the
following values taken from Ref.~\cite{PDG}:
\bqa
\begin{array}{ll}
\mw = 80.398\,\GeV,  \;\; & \;\; 
\mz = 91.1876\,\GeV, \;\; \\
\Gamma_{\ssZ} = 2.4952\,\GeV, \;\; & \; \;
\gf = 1.16637\,\times\,10^{-5}\,\GeV^{-2}.
\end{array}
\eqa
For the mass of the top quark, we have used $M_t=170.9\,\GeV$~\cite{Top};
for the width of the $W$ boson, we have chosen the value
$ \Gamma_{\ssW} = 2.093\,\GeV$, predicted by the Standard Model
with electroweak and QCD corrections at one loop.

\begin{figure}[ht]
  \begin{center}
  \includegraphics[scale=1.1]{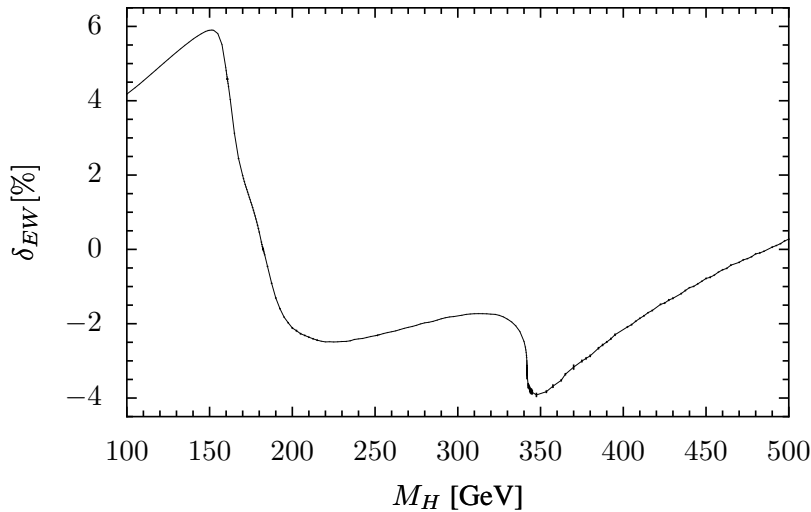}
  \caption[]{NLO electroweak percentage
    corrections to the partonic cross section
    $\sigma( g \, g \to H)$.}
     \label{fig:deltaEW}
  \end{center}
  \end{figure}

Our results for $\delta_{\EW}$ defined in \eqn{eq:deltaPART} are shown 
in \fig{fig:deltaEW}, where we include the complete corrections, 
comprehensive of light- and top-quark contributions, in the 
entire range of interest. The introduction of the complex-mass scheme in our 
two-loop evaluation has a striking consequence, visible around two-particle 
thresholds, where artificial cusp effects disappear. A detailed analysis of this
issue can be found in Ref.~\cite{newShort}.

For including the NLO electroweak corrections in
the hadronic cross section of \eqn{eq:CShad}, we have used the {\sc FORTRAN} code 
{\sc HIGGSNNLO} written by M.~Grazzini (see also Ref.~\cite{Catani:2008me}), with 
QCD corrections at NNLO, interfaced
with the {\sc MRST2002} set of parton distribution functions~\cite{Martin:2002aw}. 
Although partially outdated, they represent the best choice for our purposes,
allowing for a direct comparison with the results of Ref.~\cite{Catani:2003zt}.
In the code the parton densities, as well as the strong-coupling constant 
$\alpha_{\ssS}$, are evaluated at each corresponding order, with
one-loop $\alpha_{\ssS}$ at LO ($\alpha_{\ssS}(\mzs)=0.130$), two-loop 
$\alpha_{\ssS}$ at NLO ($\alpha_{\ssS}(\mzs)=0.1197$) and three-loop 
$\alpha_{\ssS}$ at NNLO ($\alpha_{\ssS}(\mzs)=0.1154$), as described
in Ref.~\cite{Catani:2003zt}.

In the following, we will discuss our numerical results introducing $K$ factors,
defined as the ratio of the cross section including higher order corrections
over the LO result. According to the discussion in Section~\ref{sec:inclu},
we will define $K$ factors for NNLO QCD corrections and for
NNLO QCD $+$ NLO electroweak corrections under the assumption
of complete (partial) factorization. The LO expression which normalizes each $K$ 
factor will be always evaluated for $\mu_\ssR=\mu_\ssF=\mh$, with parton 
densities and $\alpha_\ssS$ evolved at LO.
\subsection{LHC results}
\begin{figure}[ht]
    \begin{center}
      \includegraphics[scale=0.78]{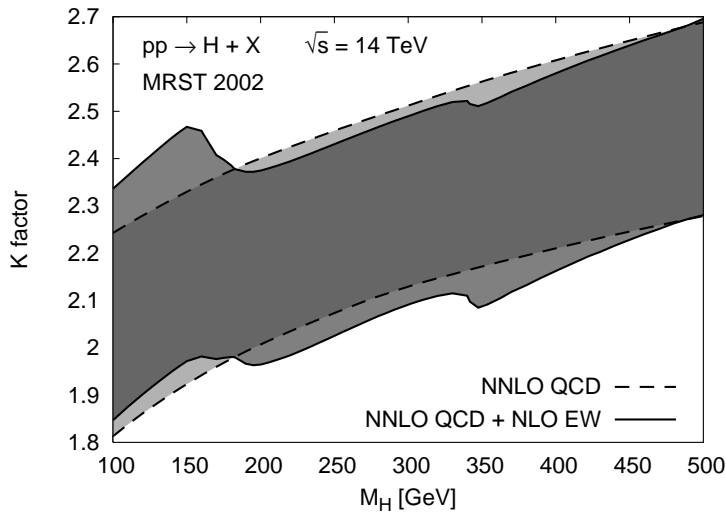}
      \vspace{-0.3cm}
      \caption[]{Uncertainty bands for the $K$ factors
        for Higgs production at the LHC.}
      \label{fig:Kfact}
    \end{center}
  \end{figure}

We start showing our results for a LHC center-of-mass energy $\sqrt{s}= 14\,$ TeV.
According to the standard procedure described in Section~\ref{sec:inclu},
an uncertainty band for the $K$ factors is derived exploring the dependence of the 
cross section on the renormalization and factorization scales. In \fig{fig:Kfact}
we show the maximal and minimal values as a function of $\mh$, for
$100\,$ GeV $\le \mh \le 500\,$ GeV, obtained including NNLO QCD corrections
only (dashed lines) and NNLO QCD $+$ NLO electroweak ones (solid lines). Note
that in the second case the maximal and minimal values take into account
both factorization options of \eqn{eq:CF} and \eqn{eq:PF}.
The pattern of the electroweak-corrected uncertainty band for the $K$ factors
in \fig{fig:Kfact} clearly follows the shape of the correction factor at the
partonic level of \fig{fig:deltaEW}. At about $180\,$ GeV, when 
$\delta_{\EW}$ becomes negative, the two factorization options exchange 
their role: below $180\,$ GeV, the completely (partially) factorized result 
fixes the upper (lower) bound, above $180\,$ GeV the completely (partially) 
factorized result fixes the lower (upper) bound. This is just a consequence
of the fact that the assumption of a complete factorization leads to
a larger enhancement of the absolute value of the NNLO QCD corrections.

In contrast with the purely QCD-corrected $K$ factors, we observe a higher
sensitivity of the electroweak-corrected shapes of \fig{fig:Kfact} on the values 
of the Higgs mass. The larger effect takes place at $\mh=150\,$ GeV, where the QCD 
band at NNLO ranges from $1.92$ to $2.33$ and NLO electroweak effects shift it 
to an interval between $1.97$ and $2.47$.

The corresponding numerical values for the NNLO QCD-corrected cross section 
$\sigma^{\rm QCD}$,
and the cross section including NNLO QCD $+$ NLO electroweak corrections
under the assumption of complete (partial) factorization, $\sigma^{\rm CF}$ 
($\sigma^{\rm PF}$), are shown in \tabn{tab:LHC}, including minimal, reference and 
maximal values (the differences with Ref.~\cite{Catani:2003zt} are simply due to 
the different values used for the top-quark mass, $M_t=170.9\,$ GeV in this paper, 
$M_t=176\,$ GeV in Ref.~\cite{Catani:2003zt}).

Strictly speaking, the values for the QCD corrections at NNLO are given
here in the large-$M_t$ limit; however, it has been shown at NLO that
this approximation is extremely good for $\mh \le 2\, M_t$
and works with a good accuracy up to $\mh=1$ TeV~\cite{Kramer:1996iq}.
\renewcommand{\arraystretch}{1.3}
\begin{table}[ht]
  \begin{center}
    \begin{tabular}{|r||r|r|r||r|r|r||r|r|r|}
      \hline
      $M_H$ & 
        $\sigma^{\rm QCD}_{\rm min} $ & $\sigma^{\rm QCD}_{\rm ref} $ & $\sigma^{\rm QCD}_{\rm max} $ 
      & $\sigma^{\rm CF }_{\rm min} $ & $\sigma^{\rm CF }_{\rm ref} $ & $\sigma^{\rm CF }_{\rm max} $
      & $\sigma^{\rm PF }_{\rm min} $ & $\sigma^{\rm PF }_{\rm ref} $ & $\sigma^{\rm PF }_{\rm max} $ \\
      \hline
      \hline
      100 & 47.84 & 53.44 & 59.18 & 49.83 & 55.68 & 61.65 & 48.73 & 54.55 & 60.57 \\ 
      110 & 41.22 & 45.92 & 50.72 & 43.09 & 48.01 & 53.02 & 42.05 & 46.94 & 51.99 \\ 
      120 & 35.94 & 39.96 & 44.03 & 37.71 & 41.92 & 46.19 & 36.72 & 40.91 & 45.21 \\ 
      130 & 31.66 & 35.13 & 38.63 & 33.34 & 36.99 & 40.68 & 32.40 & 36.02 & 39.74 \\ 
      140 & 28.14 & 31.16 & 34.20 & 29.73 & 32.92 & 36.14 & 28.83 & 31.99 & 35.24 \\ 
      150 & 25.20 & 27.86 & 30.53 & 26.69 & 29.50 & 32.33 & 25.84 & 28.63 & 31.49 \\ 
      160 & 22.73 & 25.08 & 27.45 & 23.82 & 26.29 & 28.77 & 23.19 & 25.64 & 28.15 \\ 
      170 & 20.62 & 22.73 & 24.84 & 21.03 & 23.18 & 25.33 & 20.79 & 22.94 & 25.10 \\ 
      180 & 18.82 & 20.72 & 22.61 & 18.91 & 20.81 & 22.72 & 18.86 & 20.76 & 22.67 \\ 
      190 & 17.27 & 18.99 & 20.70 & 17.04 & 18.74 & 20.43 & 17.18 & 18.87 & 20.56 \\ 
      200 & 15.93 & 17.49 & 19.05 & 15.59 & 17.12 & 18.65 & 15.79 & 17.32 & 18.85 \\ 
      220 & 13.75 & 15.07 & 16.39 & 13.41 & 14.70 & 15.98 & 13.61 & 14.91 & 16.18 \\ 
      240 & 12.11 & 13.25 & 14.38 & 11.81 & 12.93 & 14.03 & 11.99 & 13.10 & 14.21 \\ 
      260 & 10.87 & 11.87 & 12.87 & 10.63 & 11.61 & 12.58 & 10.77 & 11.76 & 12.73 \\ 
      280 &  9.96 & 10.87 & 11.76 &  9.76 & 10.65 & 11.53 &  9.88 & 10.77 & 11.65 \\ 
      310 &  9.15 &  9.98 & 10.80 &  8.99 &  9.81 & 10.61 &  9.09 &  9.91 & 10.70 \\ 
      340 &  9.38 & 10.24 & 11.07 &  9.15 &  9.98 & 10.79 &  9.29 & 10.13 & 10.93 \\ 
      370 & 10.50 & 11.46 & 12.39 & 10.16 & 11.10 & 11.99 & 10.37 & 11.31 & 12.19 \\ 
      410 &  9.00 &  9.83 & 10.62 &  8.83 &  9.65 & 10.42 &  8.94 &  9.75 & 10.52 \\ 
      450 &  6.91 &  7.55 &  8.15 &  6.86 &  7.49 &  8.09 &  6.89 &  7.52 &  8.12 \\ 
      500 &  4.73 &  5.17 &  5.58 &  4.74 &  5.18 &  5.59 &  4.73 &  5.17 &  5.59 \\ 
      \hline 
    \end{tabular}
  \end{center}
  \caption[]{NNLO QCD and NNLO QCD $+$ NLO electroweak (with 
    CF (\eqn{eq:CF}) and PF (\eqn{eq:PF}) options) cross sections in pb as a 
    function of the Higgs mass in GeV at the LHC.}
  \label{tab:LHC}
\end{table}
\subsection{Tevatron results}
\begin{figure}[ht]
    \begin{center}
      \includegraphics[scale=0.78]{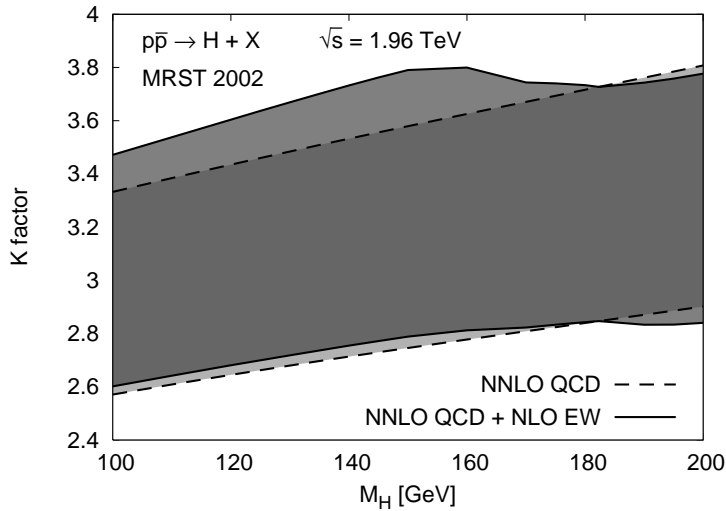}
      \caption[]{Uncertainty bands for the $K$ factors
        for Higgs production at the Tevatron.}
      \label{fig:Kfact2}
    \end{center}
  \end{figure}

In this section we briefly summarize the results for the Tevatron Run II, with
a center-of-mass energy $\sqrt{s}= 1.96$ TeV.
In \fig{fig:Kfact2} we show the maximal and minimal values for the $K$ factors
as a function of $\mh$ in the range $100\,$ GeV $\le \mh \le 200\,$ GeV, obtained 
including NNLO QCD corrections (dashed lines) and NNLO QCD $+$ NLO electroweak ones 
(solid lines). As usual, maximal and minimal values take into account both 
factorization options of \eqn{eq:CF} and \eqn{eq:PF}.
The corresponding numerical values are shown in \tabn{tab:TeV}.

As stressed in Ref.~\cite{Catani:2003zt}, the $K$ factors at the Tevatron are 
larger than those at the LHC, because the Higgs production takes place closer to 
the hadronic threshold and soft-gluon effects are extremely relevant. Consequently,
the impact of soft-gluon resummation is more sizeable than the electroweak effects:
for $\mh=120\,$ GeV, the NNLL result of Ref.~\cite{Catani:2003zt} increases the 
fixed-order NNLO value by $12 \, \%$; from \tabn{tab:TeV}, the impact of NLO 
electroweak terms amounts to $+ 5 \, \%$ ($+ 2 \, \%$) assuming a complete 
(partial) factorization.

Note that the same consideration is not true for the LHC, where NLO electroweak 
effects are comparable to those due to soft-gluon resummation at NNLL accuracy. 
For $\mh=120\,$ GeV, we get from Ref.~\cite{Catani:2003zt} an increase of 
$6 \, \%$, which matches the electroweak effects; the latter amount to 
$ + 5 \, \%$ ($+ 2 \, \%$) assuming a complete (partial) factorization (see 
\tabn{tab:LHC}). Further details will be given in Section~\ref{sec:NNLL}.
\begin{table}[ht]
  \begin{center}
    \begin{tabular}{|r||r|r|r||r|r|r||r|r|r|}
      \hline
      $M_H$ & 
        $\sigma^{\rm QCD}_{\rm min}$ & $\sigma^{\rm QCD}_{\rm ref}$ & $\sigma^{\rm QCD}_{\rm max}$ 
      & $\sigma^{\rm CF }_{\rm min}$ & $\sigma^{\rm CF }_{\rm ref}$ & $\sigma^{\rm CF }_{\rm max}$
      & $\sigma^{\rm PF }_{\rm min}$ & $\sigma^{\rm PF }_{\rm ref}$ & $\sigma^{\rm PF }_{\rm max}$ \\
      \hline
      \hline
      100 & 1.2981 & 1.4886 & 1.6829 & 1.3523 & 1.5508 & 1.7532 & 1.3136 & 1.5097 & 1.7122 \\ 
      110 & 1.0002 & 1.1482 & 1.2978 & 1.0455 & 1.2003 & 1.3567 & 1.0130 & 1.1656 & 1.3222 \\ 
      120 & 0.7833 & 0.9002 & 1.0175 & 0.8219 & 0.9445 & 1.0675 & 0.7940 & 0.9147 & 1.0380 \\ 
      130 & 0.6218 & 0.7153 & 0.8086 & 0.6548 & 0.7533 & 0.8515 & 0.6307 & 0.7276 & 0.8260 \\ 
      140 & 0.4993 & 0.5750 & 0.6500 & 0.5276 & 0.6075 & 0.6868 & 0.5068 & 0.5854 & 0.6648 \\
      150 & 0.4049 & 0.4668 & 0.5279 & 0.4288 & 0.4943 & 0.5590 & 0.4112 & 0.4755 & 0.5403 \\ 
      160 & 0.3314 & 0.3824 & 0.4326 & 0.3473 & 0.4008 & 0.4534 & 0.3355 & 0.3882 & 0.4408 \\ 
      170 & 0.2734 & 0.3158 & 0.3573 & 0.2788 & 0.3220 & 0.3644 & 0.2748 & 0.3177 & 0.3601 \\ 
      180 & 0.2272 & 0.2627 & 0.2974 & 0.2283 & 0.2639 & 0.2988 & 0.2275 & 0.2630 & 0.2979 \\ 
      190 & 0.1901 & 0.2200 & 0.2492 & 0.1876 & 0.2171 & 0.2459 & 0.1895 & 0.2191 & 0.2479 \\ 
      200 & 0.1601 & 0.1854 & 0.2101 & 0.1567 & 0.1815 & 0.2057 & 0.1593 & 0.1843 & 0.2084 \\ 
      \hline 
    \end{tabular}
  \end{center}
  \caption[]{NNLO QCD and NNLO QCD $+$ NLO electroweak (with 
    CF (\eqn{eq:CF}) and PF (\eqn{eq:PF}) options) cross sections in pb as a 
    function of the Higgs mass in GeV  at the Tevatron.}
  \label{tab:TeV}
\end{table}
\subsection{Threshold behavior}
The scheme employed in our computation~\cite{newLong,newShort}
shows that the partonic cross section $\sigma(g g \to H)$ is
a smooth function of the Higgs mass; as illustrated in \fig{fig:deltaEW},
there are no artifical large effects at the opening of two-particle thresholds.

As a consequence, unphysical cusps are avoided also at the
hadronic level; in \tabn{tab:EWexa} we summarize the results at the LHC
for values of the Higgs mass corresponding to the $WW$, $ZZ$ and
$ t \overline{t}$ thresholds, following the same 
pattern of \tabn{tab:LHC} and \tabn{tab:TeV}.
\begin{table}[ht]
  \begin{center}
    \begin{tabular}{|r||r|r|r||r|r|r||r|r|r|}
      \hline
      $M_H$ & 
        $\sigma^{\rm QCD}_{\rm min} $ & $\sigma^{\rm QCD}_{\rm ref} $ & $\sigma^{\rm QCD}_{\rm max} $ 
      & $\sigma^{\rm CF }_{\rm min} $ & $\sigma^{\rm CF }_{\rm ref} $ & $\sigma^{\rm CF }_{\rm max} $
      & $\sigma^{\rm PF }_{\rm min} $ & $\sigma^{\rm PF }_{\rm ref} $ & $\sigma^{\rm PF }_{\rm max} $ \\
      \hline
      \hline
      160.8 & 22.55 & 24.88 & 27.23 & 23.58 & 26.02 & 28.47 & 22.99 & 25.41 & 27.88 \\ 
      182.3 & 18.44 & 20.29 & 22.14 & 18.44 & 20.29 & 22.14 & 18.44 & 20.29 & 22.14 \\ 
      341.8 &  9.50 & 10.37 & 11.21 &  9.21 & 10.05 & 10.86 &  9.39 & 10.23 & 11.04 \\ 
      \hline 
    \end{tabular}
  \end{center}
  \caption[]{NNLO QCD and NNLO QCD $+$ NLO electroweak (with 
    CF (\eqn{eq:CF}) and PF (\eqn{eq:PF}) options) cross sections in pb at 
    the two-particle thresholds at the LHC. $\mh$ is given in GeV.}
\label{tab:EWexa}
\end{table}

At the $WW$ threshold, electroweak effects amount to $+5 \%$ (complete 
factorization, CF) and $+2 \%$ (partial factorization, PF) of the NNLO QCD result;
around the $ZZ$ threshold, they are vanishingly small, since here the corrections 
to the partonic cross section of \fig{fig:deltaEW} are negligible. Finally, the 
effect at the $t {\bar t}$ threshold is to moderately decrease the cross section, 
by an amount of $-3 \%$ (CF) and $-1 \%$ (PF).
\subsection{Comparison with NNLL soft-gluon resummation}
\label{sec:NNLL}
In \tabn{tab:moves} we show a comparison, for $\sqrt{s}=14$ TeV, of
our result with the NNLL resummation performed by the authors of 
Ref.~\cite{Catani:2003zt}.
For the NNLO QCD cross section (with and without NLO electroweak corrections)
we define an average value, $\sigma_{\rm aver}= ( \sigma_{\rm max} + 
\sigma_{\rm min} )/2$, and the associated error $\Delta \sigma = 
( \sigma_{\rm max} - \sigma_{\rm min} )/2$.
Maximal and minimal values have been shown in \tabn{tab:LHC};
for electroweak corrections, we have taken into account both CF and PF options. 
The resulting electroweak shift can be directly derived from the second and third 
columns of \tabn{tab:moves}; the NNLL shift is obtained from Tab.~1 of 
Ref.~\cite{Catani:2003zt}, taking the difference of the two reference values for 
the fixed-order and resummed computations.
\begin{table}[!ht]
\begin{center}
\begin{tabular}{|l||l||l|l|l|}
\hline
$\mh\,$ & $\sigma^{\rm QCD}_{\rm aver} \pm \Delta\sigma^{\rm QCD} $ 
        & $\sigma^{\rm EW}_{\rm aver}\pm \Delta\sigma^{\rm EW}$ & EW shift & NNLL shift  \\
\hline
\hline
110 & 45.97 $\pm$ 4.75  & 47.53 $\pm$ 5.49 & $+$ 1.56 & $+$ 2.64 \\
130 & 35.15 $\pm$ 3.48  & 36.54 $\pm$ 4.14 & $+$ 1.39 & $+$ 2.04 \\
150 & 27.87 $\pm$ 2.66  & 29.08 $\pm$ 3.25 & $+$ 1.21 & $+$ 1.62 \\
170 & 22.73 $\pm$ 2.11  & 23.06 $\pm$ 2.27 & $+$ 0.33 & $+$ 1.34 \\
190 & 18.99 $\pm$ 1.72  & 18.80 $\pm$ 1.76 & $-$ 0.19 & $+$ 1.11 \\
200 & 17.49 $\pm$ 1.56  & 17.22 $\pm$ 1.63 & $-$ 0.27 & $+$ 1.04 \\
220 & 15.07 $\pm$ 1.32  & 14.80 $\pm$ 1.39 & $-$ 0.27 & $+$ 0.89 \\
240 & 13.24 $\pm$ 1.14  & 13.01 $\pm$ 1.20 & $-$ 0.23 & $+$ 0.79 \\
260 & 11.87 $\pm$ 1.00  & 11.68 $\pm$ 1.05 & $-$ 0.19 & $+$ 0.72 \\
280 & 10.86 $\pm$ 0.90  & 10.70 $\pm$ 0.94 & $-$ 0.16 & $+$ 0.65 \\
\hline 
\end{tabular}
\end{center}
\caption[]{Shifts induced at the LHC on the NNLO QCD cross
section by NLO electroweak effects and NNLL resummation taken from 
Ref.~\cite{Catani:2003zt}. The Higgs mass is in GeV, all values for
the cross section and the shifts are in pb.}
\label{tab:moves}
\end{table}

We observe that for $\mh \le 160\,$ GeV the inclusion of NLO electroweak 
corrections increases the average value of the cross section by about half the 
size of the QCD uncertainty half-band; the effect is comparable with the impact 
of soft-gluon resummation at NNLL accuracy. For higher values of the Higgs mass,
NLO electroweak effects are negative, and they partially screen
the impact of soft-gluon resummation on the fixed-order NNLO QCD result.
\section{Conclusions}
In this paper we have presented the impact of the NLO electroweak
corrections to the inclusive cross section for Higgs production at the LHC. 

The central value of the NNLO QCD cross section for $\mh=120\,$ GeV
is shifted by  $+ 5\, \%$, both at the LHC and the Tevatron, under
the assumption of a complete factorization of the NLO electroweak
effects with respect to the dominant QCD radiation. The impact is
relevant in view of the estimated uncertainty associated with
uncalculated higher order QCD corrections, $5\, \%$ at the LHC
and $7\, \%$ at the Tevatron~\cite{Moch:2005ky}. The underlying
assumption is motivated by the observation that for low Higgs masses
the electroweak interaction is effectively point-like~\cite{Aglietti:2006yd}.
We have also derived a more conservative estimate on the electroweak shift
assuming a partial factorization at leading order: for $\mh=120\,$ GeV
the result is reduced to $+ 2\, \%$.

The NNLO QCD fixed-order uncertainty bands, derived varying the
renormalization and factorization scales at fixed values for
the Higgs mass, have been refined including the NLO electroweak
effects; as a result, they show a stronger sensitivity to the
Higgs mass with respect to the pure QCD result.

For low values of the Higgs mass, we get a qualitative agreement
with the results of Ref.~\cite{Aglietti:2006yd}, as a consequence
of the light-quark dominance. For higher values of the Higgs mass,
the role of the top quark becomes relevant and the agreement
starts deteriorating.

We have shown that large two-particle threshold effects are avoided working
in the complex-mass scheme of Ref.~\cite{Denner:2005fg}. Further details
can be found in our companion paper~\cite{newShort}.

We have performed a detailed analysis for a wide Higgs mass range,
up to $200\,$ GeV at the Tevatron and $500\,$ GeV at the LHC. Concerning the
Tevatron, we have confirmed the expectations of Ref.~\cite{Catani:2003zt},
showing that electroweak effects are considerably smaller than those
induced by the soft-gluon resummation. At the LHC, instead, the size of
the positive NLO electroweak corrections is comparable to that of the
positive soft-gluon resummation at NNLL; above $180\,$ GeV, they are
negative and moderately screen soft-gluon effects.

In summary, electroweak effects to Higgs production at hadron colliders
are under control at NLO in the whole Higgs mass range. The main source
of uncertainty is connected with a more precise knowledge of
the parton distribution functions.
\Acknowledgments
We gratefully thank Massimiliano Grazzini for allowing the use of the 
numerical program {\sc HIGGSNNLO} and for useful discussions.
We also acknowledge important discussions with Giuseppe Degrassi and Fabio Maltoni.

\end{document}